\documentclass{PoS}

\newcommand{\srm}[1]{\textrm{\scriptsize #1}}

\title{Dynamical Chirally Improved Quarks: First Results for Hadron Masses}

\ShortTitle{Dynamical Chirally Improved Quarks: First Results for Hadron Masses}

\author{R. Frigori,$^a$ Ch. Gattringer,$^a$ \speaker{C. B. Lang},$^a$   M. Limmer,$^a$ T. Maurer,$^b$
        D. Mohler$\,^a$\break and A. Sch\"afer$\,^b$\\
\llap{$^a$}
Institut f\"ur Physik, FB Theoretische Physik,
Karl-Franzens-Universit\"at Graz\\
A-8010 Graz, Austria\\
\llap{$^b$}
Institut f\"ur Theoretische Physik,
Universit\"at Regensburg\\
D-93040 Regensburg, Germany\\

E-mail: 
\email{rafael.frigori@uni-graz.at}, 
\email{christof.gattringer@uni-graz.at},
\email{christian.lang@uni-graz.at},
\email{markus.limmer@uni-graz.at},
\email{thilo.maurer@physik.uni-regensburg.de},
\email{daniel.mohler@uni-graz.at},
\email{andreas.schaefer@physik.uni-regensburg.de}

}

\abstract{We present first results for a study with two mass-degenerate
dynamical Chirally Improved (CI) fermions on lattices of spatial extent 2.4 fm.
The CI Dirac operator obeys the Ginsparg-Wilson condition in good approximation.
The pion mass we use is still large $\mathcal{O}$(470 MeV) for the $16^3\times 32$
lattices with lattice spacing of 0.15 fm. The hadron masses are obtained with
the variational technique and the results compared with earlier quenched
calculations with similar lattice parameters. We find indications for the
isovector, scalar meson $a_0(980)$ near the experimental value, in
contradistinction to quenched results which always produced a mass value
compatible with the first excitation $a_0(1450)$.}

\FullConference{The XXV International Symposium on Lattice Field Theory\\
		 July 30 - August 4 2007\\
		 Regensburg, Germany}

\begin{document}

\section{Introduction}

It will take some time before we will be able to compute hadron properties in
full QCD simulations with dynamical quarks that have full Ginsparg-Wilson (GW)
chiral symmetry. On the way to this aim we are working with the so-called
Chirally Improved (CI)  fermions. These are realized as truncated solutions to
the GW equations for a general ansatz for the Dirac operator
\cite{Ga01a,GaHiLa00}. For each site this fermion action includes several
hundred neighbors with distances ranging up to three links.

Extensive quenched calculations have demonstrated good  chiral as well as good
scaling behavior \cite{bgr04}. Within the Bern-Graz-Regensburg (BGR)
collaboration we have obtained results for the hadron spectrum and mesonic low
energy constants in quenched simulations. These included several lattice
spacings and volumes and the three light valence quarks. Emphasis in these
studies has been put on deriving sophisticated techniques to analyze excited
hadron states \cite{BuGaGl06a,BuGaGl06b,GaGlLa07} and we expect to utilize that
experience for full QCD configurations eventually.

The CI Dirac operator involves many terms and as such its implementation in an
HMC program is technically non-trivial. First results were obtained on
$12^3\times 24$ lattices and presented in Ref. \cite{LaMaOr05c,LaMaOr06a}.
However, the small volume (linear spatial size 1.8 fm) did not allow reliable
results for the hadron masses.  

Here we present first results for our HMC simulation on larger  $16^3\times
32$ lattices at lattice spacing $a=0.15$ fm and for two different values of the
quark masses. We use two mass degenerate light quarks, the L\"uscher-Weisz gauge
action, and stout smearing. One HMC-trajectory has $\mathcal{O}(100)$ steps for
one unit of HMC-time. More technical details may be found in \cite{LaMaOr05c}.

\begin{table}[b]
\begin{center}
\begin{tabular}{rrrrrrrr}
run & $\beta_\srm{LW}$ & $c_0$  & $n_\srm{conf}$ & $n_\srm{meas}$ & $a$ [fm]& $m_\srm{AWI}$[MeV] & $M_\pi$ [MeV]  \cr
\hline 
A & 4.65 & -0.06 & 425 & 65 & 0.153(1) & 33(1)& 473(4)\cr
B & 4.70 & -0.05 & 350 & 50 & 0.154(1) & 41(1)& 507(5) \cr
\hline
\end{tabular}
\end{center}
\caption{Parameters of the two runs discussed: run sequence, gauge coupling, 
bare mass parameter $c_0$, number of configurations $n_\srm{conf}$, number of
configurations analysed $n_\srm{meas}$, lattice spacing $a$  assuming that the
Sommer parameter $r_0=0.48$ fm, AWI-mass, pion mass.}
\label{tab1}
\end{table}

\section{Equilibration}

The approach to equilibration has been monitored by standard tests, like tracing
the configuration mean plaquette or the number of conjugate gradient steps in
the Monte-Carlo accept/reject step. For these two quantities we find
autocorrelation lengths around 2. We skip the first 100 configurations and then
analyze every fifth. Table \ref{tab1} specifies the parameters for the runs
discussed here.

\begin{figure}[t]
\begin{center}
\includegraphics*[width=8cm]{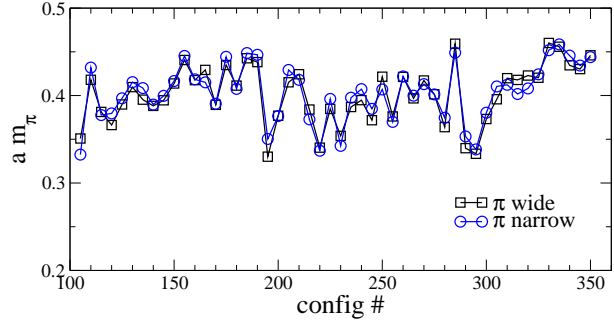}
\end{center}
\caption{\label{fig:mpi}
Time history of the pion mass determined from single configurations and
pseudoscalar operator $P=\bar d \gamma_5 u$ for run B. The quark smearing types
used (narrow for $nn$ and wide for $ww$) are indicated and show excellent
agreement.  The fits are $\cosh$-fits to the plateau-range $\Delta t = 3-15$.}
\end{figure}

On lattices of this size the determination of the pion mass is possible even for
individual configurations. Figure \ref{fig:mpi} demonstrates the (satisfactory)
time dependence of $M_\pi$ for every 5th configuration of, e.g., run B.

\begin{figure}[tp]
\begin{center}
\includegraphics*[width=6cm]{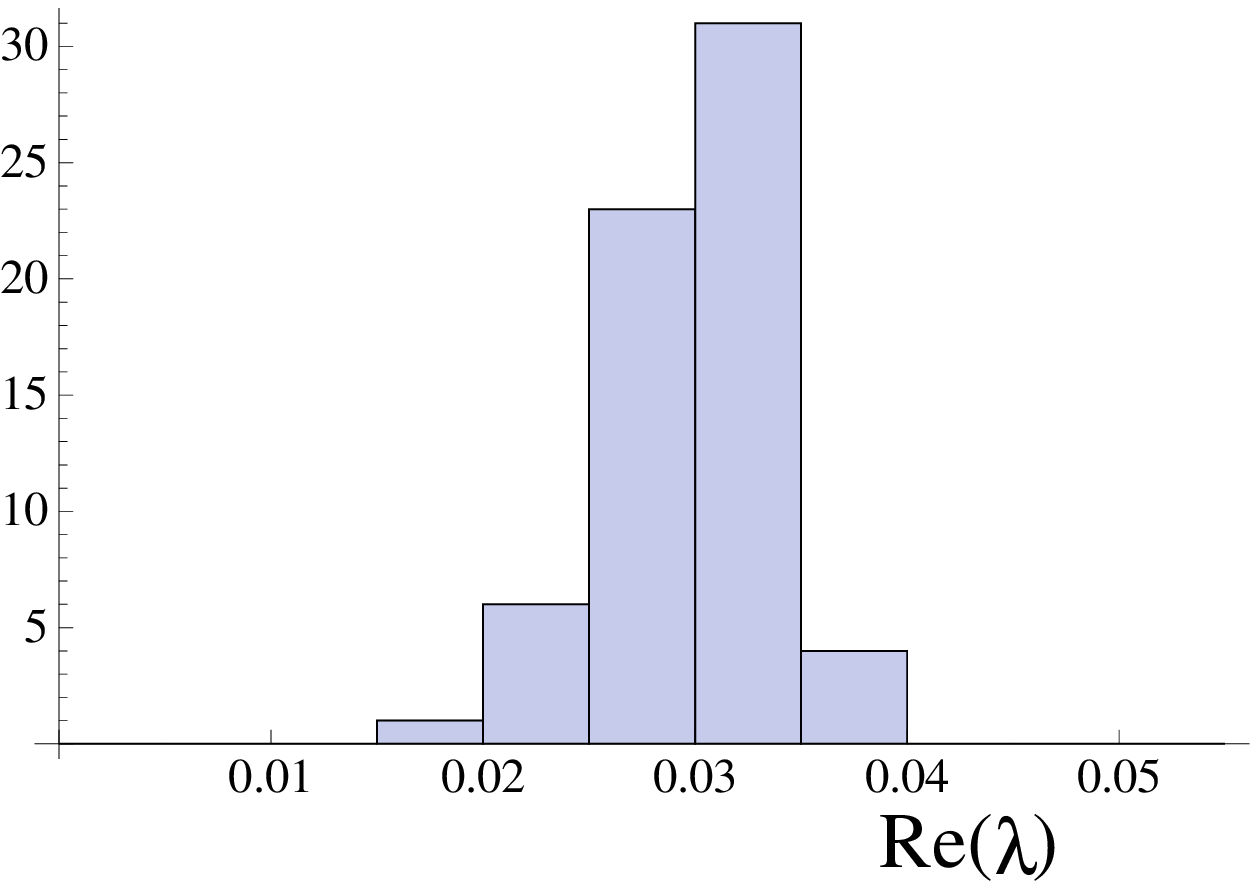}\hspace{1cm}
\includegraphics*[width=6cm]{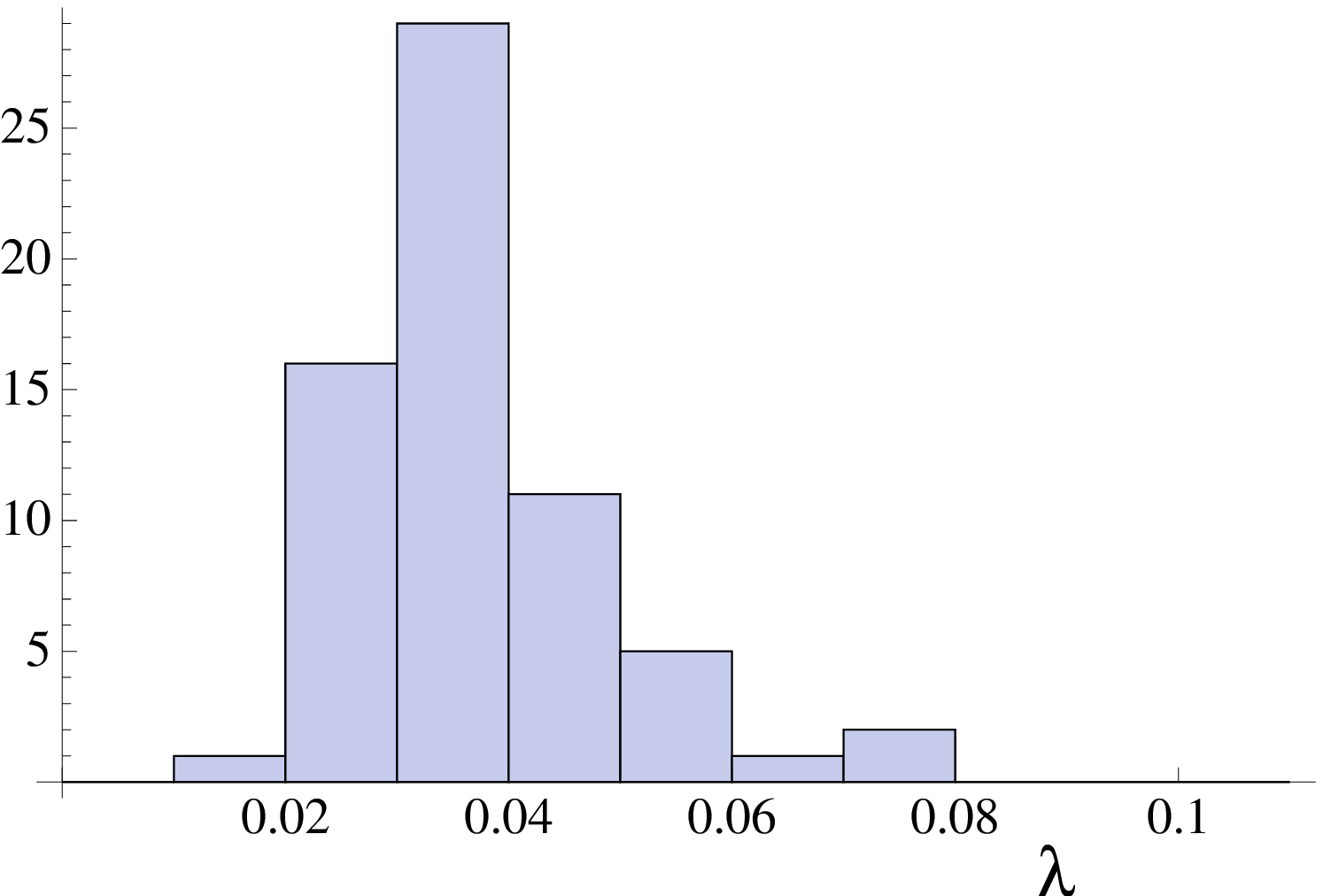}
\end{center}
\caption{\label{figs:hist}
Histogram of the smallest values of $\mathrm{Re}(\lambda)$ (left) and real
eigenvalues (right) for run A (in lattice units). The AWI-mass for this run is
at $a m_\srm{AWI}=0.0264$. }
\end{figure}

We also determined the low lying eigenvalues of the CI Dirac operator (cf.,
\cite{LaMaOr06a,JoLa07}). Figure \ref{figs:hist} shows a histogram of the
distribution of the smallest $\mathrm{Re}(\lambda)$   and of the smallest real
eigenvalues. They extend inside the GW unit circle, since the CI operator is not
exactly chiral. 

\begin{figure}[tp]
\begin{center}
\includegraphics*[width=6cm]{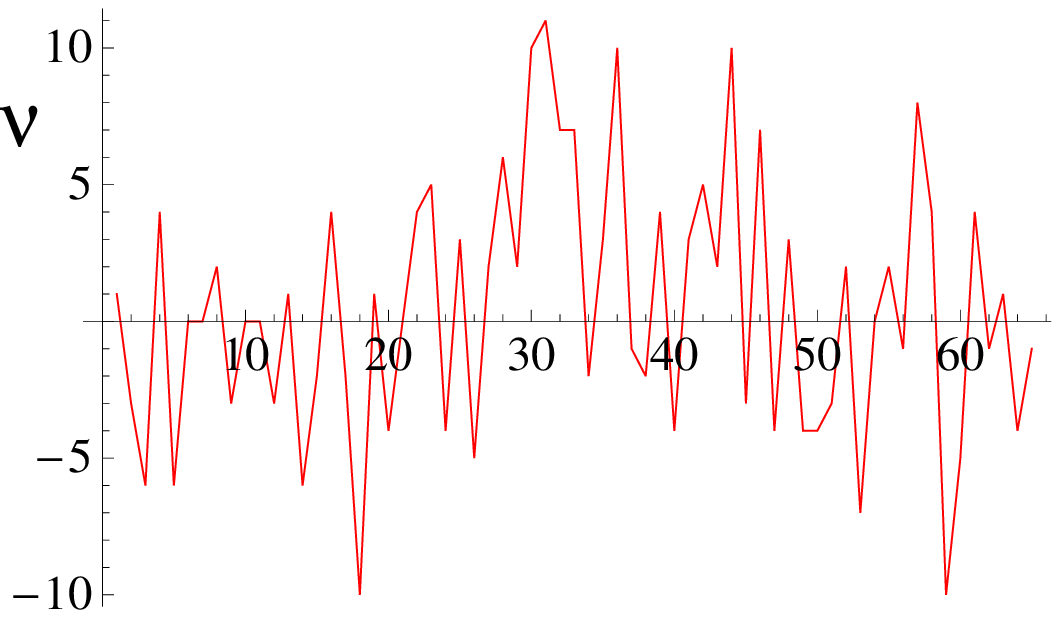}\hspace{1cm}
\includegraphics*[width=6cm]{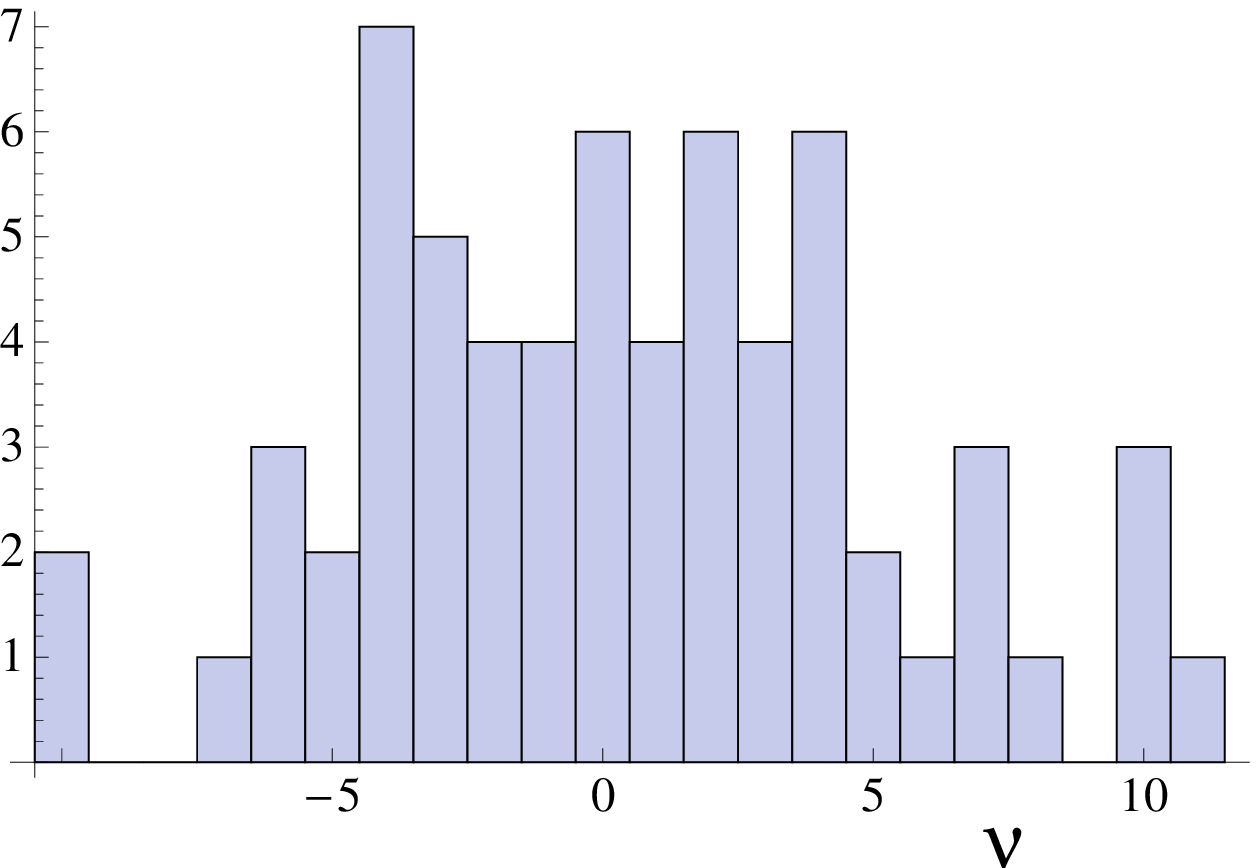}
\end{center}
\caption{\label{figs:top}
History (left) and histogram (right) of the topological charge  $\nu$ during the
run sequence A; 100 configurations have been discarded and then every 5th is
plotted.
}
\end{figure}

The topological charge $\nu$ is determined from the number of real modes counted
according to their chirality.  Tunneling is a well-known problem for the
GW-exact overlap fermions and sophisticated  techniques have been developed to
overcome this obstacle \cite{CuKrFr05,FoKaSz05}. Due to the diagonalization
method that we used (ARPACK) we may miss a few of the modes further inside the
unit GW-circle. For the CI operator we find  frequent tunneling between sectors
of different topological charge (cf., figure \ref{figs:top}).

In \cite{JoLa07} the eigenvalue distributions and properties as well as results
for a hybrid approach (overlap operator and CI configurations) are discussed in
more detail.

The axial Ward identity (AWI) allows one to define the renormalized quark  mass
through the asymptotic behavior of the ratio
\begin{equation}
\label{eq:ratDAXPX}
\frac{Z_A}{Z_P}\,
\frac{\langle \,\partial_t A_4(\vec p=\vec 0,t)\, X(0)\,\rangle}
{\langle\, P(\vec p=\vec 0,t)\,X(0) \,\rangle}
= Z_m\,2\,m=2\,m^{(r)}\;,
\end{equation}
where $X$ is any interpolator coupling to the pion and $Z_A$, $Z_P$ and $Z_m$
denote the renormalization factors relating  to the 
$\overline{\mbox{MS}}$-scheme at a scale of 2 \textrm{GeV}. These have been
calculated for the quenched case at several values of the lattice spacing and 
came out close to one \cite{GaGoHu04}. Determination of the values for the
dynamical case is under progress. For now we work with the ratio
\begin{equation}
\label{eq:ratDAPPP}
\frac{\langle \,\partial_t A_4(\vec p=\vec 0,t)\, P(0)\,\rangle}
{\langle\, P(\vec p=\vec 0,t)\,P(0) \,\rangle}
\equiv 2\,m_\srm{AWI}\;,
\end{equation}
defining the so-called AWI-mass. The effects due to smearing of the quark
sources is accounted for by comparing with point sink correlators. 

\begin{figure}[b]
\begin{center}
\includegraphics*[width=8cm]{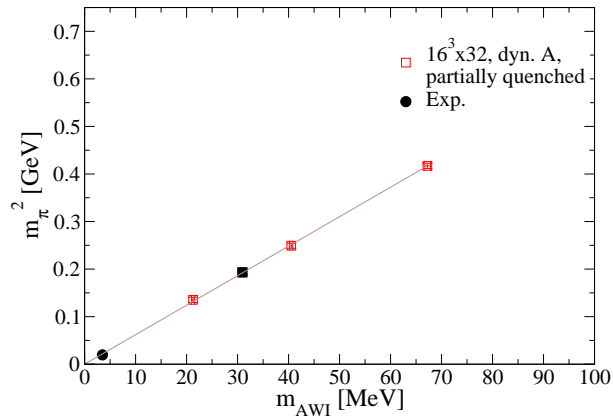}
\end{center}
\caption{\label{figs:gmor}
Pion mass squared vs. AWI-mass for run A. The data point marked
by the filled square is the value
where the valence quark and the sea quark masses agree, the other three points
are partially quenched, i.e., the valence quark masses differ from the sea quark
mass. }
\end{figure}

In figure \ref{figs:gmor} we show the pion masses obtained for run A vs.\ the
AWI-mass. In addition to the single point where $m_\srm{val}=m_\srm{sea}$ we
also exhibit three partially quenched data points, i.e., determined for
different valence quark masses for the same dynamical sea quark configurations.
We find excellent extrapolation to the chiral limit, following the leading order
(GMOR) behavior. This is no surprise since we know that the leading order
behavior is also found in purely quenched results.

\section{Hadron masses}

For the hadron propagators we applied the techniques discussed in
\cite{BuGaGl06a,BuGaGl06b}. We used Jacobi smeared quark sources and the same
set of interpolating operators for the hadrons. The correlation matrix between
the different operators at source and sink were then analyzed invoking the
so-called variational method.  The presently available statistics did not allow
to identify excited states and we therefore discuss here only results for the
ground states.

In the quenched case \cite{BuGaGl06a,BuGaGl06b} the scale has been set with 
a Sommer parameter value of $r_0=0.5$ fm. In full QCD one usually uses somewhat
smaller values. Alternative methods use $f_\pi$,  $M_\rho$ or $M_N$ to set the
scale. In most of our plots we exhibit the masses in units of the measured value
of the nucleon mass. Only for the nucleon (and for the values of the lattice
spacing in the table) we have to rely on a physical scale and there we use
$r_0=0.48$ fm.

\begin{figure}[tp]
\begin{center}
\includegraphics*[width=8cm]{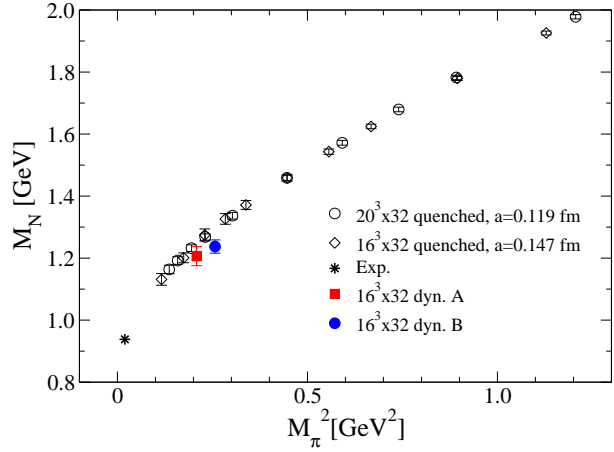}
\end{center}
\caption{\label{figs:nucleon}
Results for the ground state nucleon compared to the quenched results on a
lattice with similar parameters \cite{BuGaGl06b}.}
\end{figure}

\begin{figure}[tp]
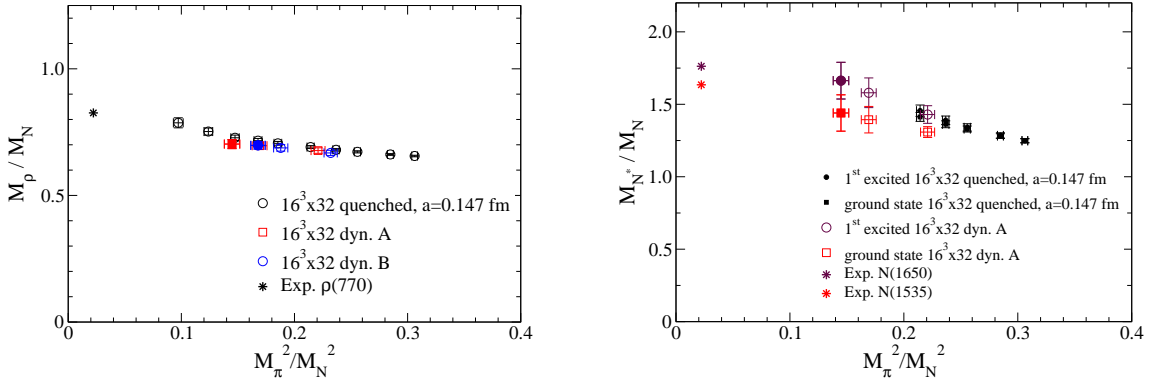

\begin{center}
\includegraphics*[width=7cm]{figs/rho_masses_APE_nucleon.eps}\hfill
\includegraphics*[width=7cm]{figs/nucleon_masses_negative_APE_nucleon.eps}
\end{center}
\caption{\label{figs:rho}
Left: Comparison of quenched (circles) and dynamical (squares) results for the
 $\rho$-meson mass in units of the measured nucleon mass. The filled squares
 denote fully dynamical results and the open ones  partially quenched results,
 both for runs A and B. Right: We can identify the two lowest lying negative
 parity nucleons which extrapolate to the experimental values.}
\end{figure}

Figure \ref{figs:nucleon} compares the results for the dynamical runs with
quenched results obtained on lattices with approximately the same physical size
and lattice spacing \cite{BuGaGl06b}. In figure \ref{figs:rho} we present a
comparison for the vector meson mass and for the negative parity nucleons; for
the latter we can identify the lowest two states.  In all cases the dynamical
results lie slightly below the quenched ones.

\begin{figure}[tp]
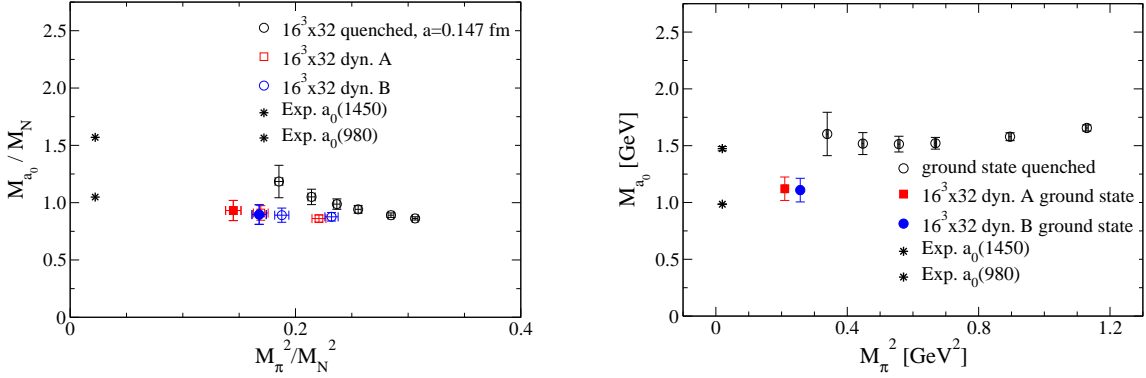

\begin{center}
\includegraphics*[width=7cm]{figs/scalar_masses_APE_nucleon.eps}\hfill
\includegraphics*[width=7cm]{figs/scalar_masses.eps}
\end{center}
\caption{\label{figs:a0}
Contrary to the quenched runs we see in the isovector, scalar channel a low
lying state compatible with the ground state $a_0$.}
\end{figure}

As a surprise came the behavior of the isovector, scalar meson $a_0$. In
quenched calculations this state was always seen  too high, compatible with the
first excitation $a_0$(1450). In the dynamical runs we now see the lowest mass 
compatible with the ground state $a_0$(980) (see figure \ref{figs:a0}). This
channel is coupled to $\pi\eta$ and we will have to study its momentum, volume
and quark mass dependence for clear identification as a single particle state.

\section{Outlook}

We are currently extending our statistics for run A and plan to also produce 
gauge configurations at significantly smaller pion masses, for similar lattice
parameters. This should allow us to find at least the first excitations for the
hadron states discussed. In particular the Roper and the Delta state are of high
interest. Also the $a_0$ signal ought to be verified by further analysis. 

\acknowledgments

The dynamical configurations have been generated on the SGI Altix 4700 of the
Leibniz Rechenzentrum Munich; the hadron propagators have then been evaluated at
the Boltzmann-Cluster at ZID at University of Graz. We thank both institutions
for providing support. Three of the authors (R.F., D.M., M.L.) are supported by
the DK W1203-N08 of the ``Fonds zur F\"orderung der wissenschaftlichen Forschung
in \"Osterreich''. A.S. and T.M. acknowledge support by DFG and BMBF.

\providecommand{\href}[2]{#2}
\begingroup\raggedright

\endgroup
\end{document}